\documentclass[usenatbib,useAMS,letterpaper]{mn2e}
\usepackage{float}
\usepackage{graphicx}
\usepackage{fixltx2e}
\usepackage{ textcomp }
\usepackage{epstopdf}
\usepackage{ctable}
\usepackage{url}
\usepackage{times}
\usepackage{amsmath}
\usepackage{amssymb}
\usepackage{xspace}
\usepackage{mathrsfs} 
\usepackage{tabularx}
\usepackage{color}

\newcommand{\msun}{M_{\odot}}

\newcommand{\acknowledgments}{\begin{small}\section*{Acknowledgements}\end{small}}

\newcommand\sref[1]{\hyperref[#1]{\S~\ref*{#1}}}
\newcommand\fref[1]{\hyperref[#1]{Fig.~\ref*{#1}}}
\newcommand\Eqref[1]{equation~(\hyperref[#1]{\ref*{#1}})}
\newcommand\tref[1]{\hyperref[#1]{Table~\ref*{#1}}}
\newcommand\aref[1]{\hyperref[#1]{Appendix~\ref*{#1}}}

\newcommand{\ksu}[1]{\textcolor{black}{#1}}

\title[Discrete Effects in Stellar Feedback]{Discrete Effects in Stellar Feedback: Individual Supernovae, Hypernovae, and IMF Sampling in Dwarf Galaxies}
\author[K. Su et al.]{
\parbox[t]{\textwidth}{
Kung-Yi Su$^{1}$\thanks{E-mail: ksu@caltech.edu}, Philip F. Hopkins$^{1}$, Christopher C. Hayward$^{2,3}$, Xiangcheng Ma$^{1}$, Michael Boylan-Kolchin$^4$, Daniel Kasen$^{5,6}$, %Eliot Quataert$^5$, 
Du\v san Kere\v s$^7$,
Claude-Andr\'e Faucher-Gigu\`ere$^8$,   Matthew E. Orr$^{1}$
}
\vspace*{6pt} \\
$^1$TAPIR 350-17, California Institute of Technology, 1200 E. California Boulevard, Pasadena, CA 91125, USA\\
$^2$Center for Computational Astrophysics, Flatiron Institute, 162 Fifth Avenue, New York, NY 10010, USA\\
$^3$Harvard-Smithsonian Center for Astrophysics, 60 Garden Street, Cambridge, MA 02138, USA\\
$^4$Department of Astronomy, The University of Texas at Austin, 2515 Speedway, Stop C1400, Austin, TX 78712, USA\\
$^5$Department of Astronomy and Theoretical Astrophysics Center, University of California Berkeley, Berkeley, CA 94720, USA\\
$^6$Lawrence Berkeley National Laboratory, 1 Cyclotron Road, Berkeley, CA 94720, USA\\
$^7$Department of Physics, Center for Astrophysics and Space Sciences, University of California at San Diego, 9500 Gilman Drive, La Jolla, CA 92093, USA\\
$^8$Department of Physics and Astronomy and CIERA, Northwestern University, 2145 Sheridan Road, Evanston, IL 60208, USA
 }

\usepackage[breaklinks,colorlinks,citecolor=blue]{hyperref}
\usepackage[all]{hypcap}

\begin{document}
\long\def\/*#1*/{}
%\date{Accepted ???. Received ???; in original form ???}
\date{Submitted to MNRAS}

\pagerange{\pageref{firstpage}--\pageref{lastpage}} \pubyear{2016}

\maketitle

\label{firstpage}

\begin{abstract}
Using high-resolution simulations from the FIRE-2 (Feedback In Realistic Environments) project, we study the effects of discreteness in stellar feedback processes on the evolution of galaxies and the properties of the interstellar medium (ISM). We specifically consider the discretization of supernovae (SNe), including hypernovae (HNe), and sampling the initial mass function (IMF). We study these processes in cosmological simulations of dwarf galaxies with $z=0$ stellar masses $M_{\ast}\sim 10^{4}-3\times10^{6}\,\msun$ (halo masses $\sim 10^{9}-10^{10}\,\msun$). We show that the discrete nature of individual SNe (as opposed to a model in which their energy/momentum deposition is continuous over time, similar to stellar winds) is crucial in generating a reasonable ISM structure and galactic winds and in regulating dwarf stellar masses. However, once SNe are discretized, accounting for the effects of IMF sampling on continuous mechanisms such as radiative feedback and stellar mass-loss (as opposed to adopting IMF-averaged rates) has weak effects on galaxy-scale properties. We also consider the effects of rare HNe events with energies $\sim 10^{53}\,{\rm erg}$. The effects of HNe are similar to the effects of clustered explosions of SNe -- which are already captured in our default simulation setup -- and do not quench star formation (provided that the HNe do not dominate the total SNe energy budget), which suggests that HNe yield products should be observable in ultra-faint dwarfs today.
\end{abstract}

\begin{keywords}
methods: numerical --- galaxies: star formation --- supernovae: general   --- ISM: structure ---  ISM: jets and outflows ---cosmology:theory
\end{keywords}

%\vspace{-0.5cm}
%\vspace{1.5cm}

\section{Introduction} \label{S:intro}

Stellar feedback is crucial in galaxy evolution. Without feedback, gas cools onto galaxies efficiently and experiences runaway collapse and star formation, resulting in predicted stellar masses orders of magnitude higher than observed \citep{1996ApJS..105...19K,1999MNRAS.310.1087S,2000MNRAS.319..168C,2003MNRAS.339..289S,2009MNRAS.396.2332K,2010MNRAS.409.1088B,2011MNRAS.413.2935D,2011IAUS..270..235H,2011ApJ...740...74K,2011ApJ...730...11T,2011MNRAS.417..950H}. %This is corroborated by the observed Kennicutt-Schmidt (KS)  relation, which shows the gas in  galaxies is consumed in  $\sim 50 -100$  dynamical times \citep{1998ApJ...498..541K,1974ApJ...192L.149Z,1997ApJ...476..166W,1999ARA&A..37..311E,2009ApJS..181..321E} and only a few percent of gas becoming stars.

Rapid progress has been made in the last decade in modeling stellar feedback in galaxy simulations \citep[see e.g.][]{2007MNRAS.374.1479G,2009ApJ...695..292C,2012MNRAS.423.2374U,2011MNRAS.417..950H, 2012MNRAS.421.3488H,2012MNRAS.421.3522H,2015MNRAS.454.2691M,2015arXiv150900853A,2015arXiv151005644H,2014ApJ...786...64K}. In \cite{2011MNRAS.417..950H, 2012MNRAS.421.3488H}, for example, a detailed feedback model including radiation pressure, stellar winds, supernovae, and photo-heating was developed and applied to idealized isolated galaxy simulations. It was shown that this stellar feedback model was able to maintain a self-regulated multi-phase interstellar medium (ISM), with giant molecular clouds (GMCs) turning only a few percent of their mass into stars in a dynamical time, and SFRs in agreement  with observations \citep{hopkins:stellar.fb.winds,hopkins:stellar.fb.mergers,hopkins:2013.merger.sb.fb.winds}. With numerical improvements and additional cooling physics, similar models were applied to cosmological zoom-in simulations in the FIRE\footnote{Project web site: http://fire.northwestern.edu} (Feedback In Realistic Environments) project \citep{2014MNRAS.445..581H}. Subsequent work showed these feedback models could reproduce a wide range of observations, including star formation histories \citep{2014MNRAS.445..581H}, the Kennicutt-Schmidt relation \citep{2017arXiv170101788O}, the star forming ``main sequence'' and time-variability of star formation \citep[][]{2017MNRAS.466...88S}, galactic winds \citep[e.g.,][]{2015MNRAS.454.2691M, 2017MNRAS.468.4170M, 2017MNRAS.470.4698A}, the dense HI content of galaxy halos \citep{2015MNRAS.449..987F,2016MNRAS.461L..32F,2017MNRAS.469.2292H}, the implied photon escape fractions of high-redshift galaxies \citep{2016MNRAS.459.3614M},
and galaxy metallicities \citep[]{2015arXiv150402097M}. 

However, there are several properties of discrete feedback processes that without proper modeling could potentially yield very different or even unreasonable ISM phase structures and galaxy morphologies. Supernovae (SNe) are naturally discrete events and tend to be clustered in time and space. Idealized studies of the ISM have shown that if the same total amount of energy is injected continuously into the ISM rather than in discrete SNe (or at too low resolution), the energy could be effectively smeared throughout the whole galaxy and be radiated away too efficiently \citep{2015ApJ...802...99K, 2015MNRAS.450..504M,2016MNRAS.459.2311M}, thus making SNe feedback much less effective than when the spatiotemporal clustering of SNe is properly modeled \citep[e.g.,][]{2016MNRAS.456.3432G, 2017MNRAS.470L..39F}. In many simulations (including those referenced above), SNe are indeed correctly treated as individual discrete events, but this is not always the case in the literature. 

Moreover, it is common in galaxy-scale simulations to treat continuous quantities (e.g.\ stellar mass-loss and radiative heating rates) as IMF-averaged. In reality, these rates are highly variable star-to-star, with most of the feedback from OB-winds, ionizing photons, and radiation pressure coming from massive O stars. When galaxies are sufficiently massive, these effects should average out, but in dwarfs, in particular, failure to account for these fluctuations could lead to biased predictions for galaxy properties. This is certainly the case for measurements of e.g.\ the ionizing flux and spectral shapes of such systems \citep[see][]{refA,refB}. IMF sampling  gets more important when the mass resolution increases, and the baryonic particle mass fall below $\sim10^4 M_\odot$ \citep{hensler_steyrleithner_recchi_2016}. In such case, the IMF is poorly sampled in a single star particle.

In addition to the aforementioned effects, hypernovae (HNe) may be yet another important discrete feedback channel. HNe are core-collapse SNe that have energies that exceed the typical SN energy ($\sim10^{51}$ erg) by a factor of 10 or more ($E>10^{52}$ erg; \citealt{2004ASSL..302..277N,2004ApJ...607L..17P}).  Such extreme events could potentially blow out all the gas in a dwarf galaxy, consequently completely quenching star formation if the galaxy's dark matter halo is too low-mass to accrete further gas post-reionisation. Whether or not an HNe quenches star formation determines whether its yield products can be incorporated into next-generation stars, which in turn determines whether or not the yield products of HNe should be observable.

In this paper, we investigate the effects of the discretization of SNe, IMF sampling and  the inclusion of HNe on the formation of dwarf galaxies.
In \sref{S:methods}, we describe the simulations. Then, we analyze the effects on the star formation histories, morphologies, phase structures,  outflows and ionizing photon escape fractions of our simulated galaxies in \sref{S:results}. In \sref{S:discussion}, we discuss our results, and our conclusions are presented in \sref{S:conclusions}.

\section{Methodology} \label{S:methods}
The simulations use {\sc GIZMO} \citep{2015MNRAS.450...53H}\footnote{A public version of this code is available at \href{http://www.tapir.caltech.edu/~phopkins/Site/GIZMO.html}{\textit{http://www.tapir.caltech.edu/$\sim$phopkins/Site/GIZMO.html}}.}, a mesh-free, Lagrangian finite-volume Godunov-type code designed to capture both the advantages of grid-based and smoothed-particle hydrodynamics (SPH) methods, in its meshless finite mass (MFM) mode. The numerical details and tests of the method are discussed in \cite{2015MNRAS.450...53H}. The default simulations use the FIRE-2 version of the code, which is described in detail in \cite{2017arXiv170206148H}. % Cooling is followed from $10^{10}$K to $10$K, including the effects of photo-electric and photo-ionization heating and collisional, Compton, fine structure, recombination, atomic, and molecular cooling. 
 Cooling is followed from $10-10^{10}$ K, including free-free, inverse Compton, atomic, and molecular cooling, accounting for photo-ionization and photo-electric heating by a UV background \citep{2009ApJ...703.1416F} and local sources.
Star formation occurs only in molecular, self-shielding, and self-gravitating \citep{2013MNRAS.432.2647H} gas above a minimum density $n>1000$cm$^{-3}$.

\begin{table*}
\begin{center}
 \caption{Galaxy simulations}
 \label{tab:ic}
 \begin{tabular*}{\textwidth}{@{\extracolsep{\fill}}cccccccccccccccc}
% \begin{tabular}{@{}l@{}l@{}l@{}l@{}l@{}l@{}l@{}l@{}l@{}l@{}l@{}l@{}l@{\:}l@{\:}l@{}l}
 %\centered
 \hline
\hline
Simulation  &$M^{\rm vir}_{\rm halo}$  &  $R_{\rm vir}$ &$M_g$               &$M_*$               &$m_{i,1000}$            &$\epsilon^{\rm min}_{\rm gas}$ &Description         \\               
Name                     & [$M_\odot$]                        &  [kpc]                & [$M_\odot$]  & [$M_\odot$] &[1000$M_\odot$]  &[pc]       \\              
\hline
m10q        & 8.0e9                                      &   52.4                 & 8.4e6                     &1.8e6                 &0.25                            &0.52      & isolated dwarf, early-forming halo\\              
m10v         & 8.3e9                                      &   53.1                & 2.1e7                     &1.0e5                  &0.25                           &0.73    & isolated dwarf, late-forming halo \\
m09           &2.4e9                                        & 35.6                & 1.2e5                     &9.4e3                 &0.25                            &1.1       &early-forming, ultra-faint field dwarf \\

\hline 
\hline
\end{tabular*}
\end{center}
\begin{flushleft}
Parameters of the galaxy models studied here:\\
(1) Simulation name: Consistent with \cite{2017arXiv170206148H}.\\
(2) $M^{\rm vir}_{\rm halo}$: Virial mass \citep{1998ApJ...495...80B} of the main halo at $z=0$.\\
(3) $R_{\rm vir}$: Viral radius of the main halo at $z=0$.\\
(4) $M_g$: Total gas mass within $\sim 0.1 R_{vir}$ at $z=0$ ($z=2$ for m09).\\
(5) $M_*$ : Total stellar mass within $\sim 0.1 R_{vir}$ at $z=0$.\\
(6) $m_{i,1000}$: Baryonic (star and gas) mass resolution in units of 1000 $M_\odot$. Dark matter particles are always $\sim 5$ times heavier.\\
(7) $\epsilon^{\rm min}_{\rm gas}$: Minimum gravitational force softening reached by the gas in the simulation (force softenings are adaptive following the inter-particle separation). Force from a particle is exactly Keplerian at $> 1.95\epsilon_{\rm gas}$; the ``Plummer-equivalent'' softening is $\approx 0.7 \epsilon_{\rm gas}$.\\

\end{flushleft}
\end{table*}
We focus on low-mass dwarf galaxies, where the effects we explore should be more significant than in more massive galaxies. Three fully cosmological zoom-in simulations from the FIRE-2 suite \citep{2017arXiv170206148H} are included in this study: m10q (an early-forming  $10^{10}M_\odot$ halo), m10v (a late-forming $10^{10}M_\odot$ halo) and m09 (a $10^{9}M_\odot$ halo). %They are briefly summarized in \tref{tab:ic}. 
Note that the tabulated halo masses are from $z=0$.

Most of the simulations have been re-run at different resolutions, with the initial gas particle masses differing by a factor of $\sim 100$. We find all of the  conclusions of this paper are insensitive to mass resolution, so we only focus on our highest-resolution comparison suite, with properties listed in \tref{tab:ic}. For all runs, a flat $\Lambda$CDM cosmology with $h=0.702$, $\Omega_M=1-\Omega_\Lambda=0.27$, and $\Omega_b=0.046$ is adopted.

For each galaxy, we consider four variations of the stellar feedback implementation in the simulations:

\begin{enumerate}
\item{\bf Default FIRE-2 Feedback Physics (``Default''):} This is our standard FIRE-2 implementation \citep{2017arXiv170206148H}.  To summarize: once formed, a star particle is treated as a single-age stellar population with metallicity inherited from its parent gas particle and age appropriate for its formation time. All corresponding stellar feedback inputs (SNe and mass-loss rates, spectra, etc.) are determined by using {\small STARBURST99} \citep{1999ApJS..123....3L} to compute the IMF-averaged rate for a \citet{2002Sci...295...82K} IMF. The stellar feedback model includes the following: (1) radiative feedback in the form of photo-ionization and photo-electric heating, in addition to single and multiple-scattering radiation pressure with five bands (ionizing, FUV, NUV, optical-NIR, IR) tracked; (2) stellar mass loss with continuously injected mass, metals, energy, and momentum from OB and AGB winds; (3) SNe Types II and Ia using tabulated SNe rates as a function of stellar age the IMF to determine the probability of an SN originating in the star particle during each timestep\footnote{For particle masses $\approx250\,\msun$ and typical timesteps in dense star-forming gas of $\sim 100\,$yr, the probability of a SN in a young ($\sim 3-10\,$Myr old) star particle in one timestep is $dp \sim 10^{-5}$.} and then determine stochastically whether an SN occurs by drawing from a binomial distribution. If an event occurs, the appropriate gas mass, metal mass, momentum, and energy are injected -- in other words, SNe are discrete events. We assume that each SNe has an initial ejecta energy of $10^{51}\,{\rm erg}$ (see \citealt{2017arXiv170206148H,2017arXiv170707010H} for details regarding how this is coupled). \ksu{To separate the effects of IMF sampling and HNe from purely simulation stochastic effects (which vary from simulation to simulation, for the same physics), two m10q simulations are evolved with the same default physics but different random number seeds. They are labeled ``Default'' and ``Default 2,'' respectively.}

\item{\bf Continuous SNe Energy Injection (``Continuous''):} Here we take our ``Default'' model but modify it by treating SN feedback as a continuous rather than discrete process. Specifically, for each star particle, we take the expectation value for the probability of an SN occurring in a given timestep in a star particle and simply inject that {\em fraction} of a single SN's feedback-related quantities (e.g. gas mass, metal mass, energy, and momentum).\footnote{This can be as little as $\sim 10^{46}$ erg per time step in dense, star-forming gas. } Thus, the energy in this case is ``smeared'' in both time and space, as if SN feedback were continuous (as stellar winds and radiation are). The Continuous feedback simulations are not evolved all the way to $z=0$, as they become very expensive as gas catastrophically collapses into dense structures. 

\item{\bf (Approximate) IMF-Sampling Effects (``IMF-SMP''):} In this case, we take our ``Default'' model and implement a very simple approximation for the effects of discreteness resulting from IMF sampling, particularly for the radiative feedback and stellar mass-loss channels. Since the simulations are still far too low-resolution to actually resolve the IMF and the feedback channels of interest are completely dominated by massive stars, we implement an intentionally simplified ``toy model'' for IMF sampling. Specifically, each time a star particle forms, we determine the number of massive ``O stars'', $N_{\text{O}}$, from a Poisson distribution with expectation value $\langle N_{\text{O}} \rangle \approx m_{\rm particle}/100\,\msun$. All feedback rates that depend on massive stars (photo-ionization and photo-electric heating, radiation pressure in the UV, OB winds, and core-collapse SNe rates) are then scaled by the ``O-star number,'' i.e.\ their IMF-averaged rates are multiplied by $N_{O}/\langle N_{O} \rangle$ (so, by definition, the IMF-averaged rates are recovered). In the SNe case, whether SN event happen is then determine stochastically by drawing from a binomial distribution according to the updated SNe rate. Each time a core-collapse SN occurs, we delete one ``O star.'' 

\item{\bf Hypernovae (``IMF-SMP+HNe''):} Observationally, HNe are rare. One category of events that is referred to as HNe is energetic SNe associated with gamma-ray bursts (broad-lined Type Ic SN). They occur at a rate that is only $\sim 5\%$ of the Type Ib/Ic rate, with more energetic events ($E_{\rm HNe}\gtrsim 10^{52}\,{\rm erg}$) representing roughly $\sim1\%$ of the total core-collapse SNe rate \citep{2004ApJ...607L..17P,2006Natur.442.1014S,2007ApJ...657L..73G}. Another class of HNe have been theorized to come from the pair-instability SN from massive stars with  $~10^{53}{\rm erg}$ but $<10^{-4}$ of the SN rate \citep{2012Sci...337..927G}.

Here, we are interested in the most extreme events (which would have the most dramatic effects on their host galaxies), so based on the event rate distribution in \citet{1999ApJ...512L.117H}, we assume an HN energy of $E_{\rm HNe}=10^{53}\,{\rm erg}$ (i.e.\ $100\times$ a typical SN) and event rate that is $10^{-3}$ times the normal core-collapse SN rate. \footnote{This may be close to an upper limit unless the IMF is more top-heavy.} In our m10q simulation, we simply assign each core-collapse event a random probability of being an HN equal to $0.1\%$, and, if the event is defined a HNe, we increase the energy of the ejecta by a factor of $100$, but the ejecta mass is kept the same. In our m09 and m10v simulations, the stellar mass is sufficiently low that the expectation value of the number of HNe is $\lesssim1$, so we take our ``IMF-SMP'' runs, re-start them just after one of the peak star formation events (at $z=0.31$ for m10v and $z=4.0$ for m09), and manually insert a single HN explosion at that time. Note that these choices ensure that the {\em total} energy contributed by HNe is only $\sim10\%$ of the SNe budget, so we are not changing the IMF-averaged properties significantly.
\end{enumerate}

\section{Results} \label{S:results}

\subsection{Star formation rates}
\begin{figure*}
\centering
\includegraphics[width=18cm]{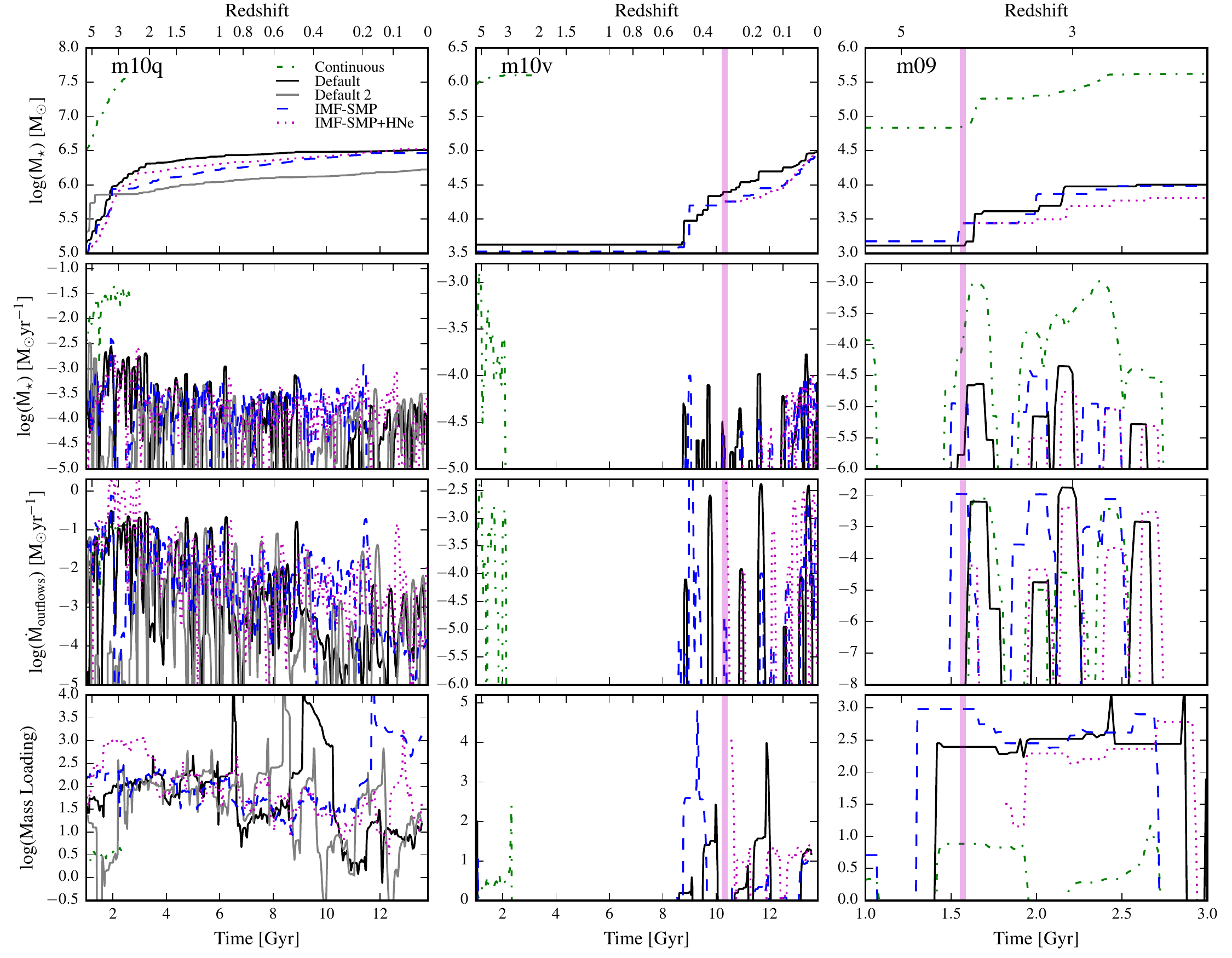}
\caption{ {\bf Top row:} Stellar mass as a function of cosmic time in our simulations. The vertical magenta lines label the times when HNe are manually exploded in the m10v and m09 runs (m10q, being more massive, has $\sim 30$ HNe randomly distributed among the SNe over its history).  {\bf Second row:} SFR averaged over the preceding 100 Myr as a function of time. {\bf Third row:} The mass outflow rate as a function of time smoothed over 100 Myr. To estimate the mass outflow rate, we consider all gas particles between 0.08 and 0.1 $r_{vir}$ that have radial velocities greater than 30 km s$^{-1}$. {\bf Bottom row:} Outflow mass-loading factor, $\eta\equiv\dot{M}_{\rm outflow}/\dot{M}_{\rm SFR}$, smoothed over 500 Myr. Treating SN feedback as continuous results in higher SFRs -- and thus stellar masses -- and lower outflow mass loading factors. 
\ksu{The final stellar mass of m10q ``Default'' and ``Default 2'' runs differ by a factor of $\sim 2$. Given such range of stochastic effect, the effect of IMF sampling or HNe is not obvious.}
%\ksu{IMF sampling results in a slight boost in the SFR in the m10q runs relative to the ``Default'' model, but this difference appears to be stochastic rather than an effect of IMF sampling since the difference is not visible in the extra ``Default'' (gray) and ``IMF-SMP'' (cyan) runs.} 
In the m09 run in which an HN was included, the final stellar mass is reduced by $\sim 0.2$ dex.
% IMF sampling in m10q boosts the outflow rate, but not the  mass-loading, suggesting that the feedback efficiency is not significantly different on the average. 
All panels are plotted after the 1st Gyr of the simulation when the halos are slightly more settled and the outflows are more well-defined.
%\cch{Perhaps add redshift axes on top of each column?}
}
\label{fig:sfr}
\end{figure*}

The first two rows of \fref{fig:sfr} show the cumulative stellar mass and SFR averaged in a 100-Myr interval for each galaxy. In all cases, the ``Continuous'' runs have an order-of-magnitude greater final stellar mass, indicating that the SN feedback is effectively weaker than in the ``Default'' model. Although the same amount of SNe energy is deposited into the surrounding gas particles in an integral sense, it is radiated away before doing significant work on the surrounding dense ISM significantly because the feedback is temporally diluted (a manifestation of the well-known overcooling problem in galaxy formation simulations). 

\ksu{On the other hand, IMF sampling does not appear to have a significant systematic effect on stellar masses, i.e. the effects of IMF sampling appear smaller than purely stochastic simulation variations. The m10q ``Default'' and ``Default 2'' runs differ significantly in star formation histories, with final stellar masses differing by a factor of $\sim 2$, even though these two runs use exactly the same physics. 
Two more m10q `IMF-SMP'' runs evolved to $z\sim 0.6$ show a similar range of stochastic differences.
%However, this appears to be a stochastic effect, as no systematic difference in stellar mass is in the galaxies that formed fewer stars (m10v and m09). To confirm the stochastic origin of the higher SFR in m10q, we simulate two more ``IMF-SMP'' runs (cyan) and one more ``Default'' run (gray) to z ∼ 1.6 and z ∼ 0.6 respectively that have different random seeds  but are otherwise identical. 
%Given these purely stochastic simulation variations, 
We thus find that the purely stochastic run-to-run variation with the same physics but with different random number seeds (resulting in variations in the detailed ages and relative positions of star particles, and therefore, the feedback injection sites) is larger than the variation when IMF sampling is included. 
The difference in SFRs among m10q runs is connected to the variations in gas phase structure and outflows, which will be discussed in \sref{S:phase} and \sref{S:outflow}.} 

In the m10q ``IMF-SMP'' run, an extreme but apparently stochastic overlap of many SNe at the same time (at $z\sim 0.2$) expels a large fraction of the galaxy's gas supply, causing a decrease in the SFR for an extended period of time. A similar event can be observed in the m10q ``IMF-SMP+HNe'' run at $z\sim 0.09$, although it is not as dramatic. These events are also a result of stochastic variations  instead of differences in the feedback implementations. %\cch{Should we rerun one of the IFM sampling runs with a different seed to demonstrate that this is \emph{not} an effect of the IMF sampling?} 
Of course, the very fact that stochastic effects can be this dramatic in  such small dwarfs owes to the fact that just a relatively small number of highly-clustered SNe can significantly perturb the galaxy.

After manually exploding HNe in m10v and m09, star formation ceases for only a few million years. HNe do not indefinitely quench star formation even in our smallest halo in this study (m09), nor do they affect the star formation histories in a qualitatively different manner from overlapping SNe events that occur after, e.g. the formation of a modest-size star cluster in a massive GMC. Note that m09 is quenched after reionisation, although it takes until $z\sim 3$ for the galaxy to exhaust its existing cold gas supply (see \citealt{2017MNRAS.471.3547F}); this behaviour is the same for all of the m09 runs considered here.

\subsection{Phase structure}
\label{S:phase}
\begin{figure*}
\begin{flushleft}
\centering
\includegraphics[width=18cm]{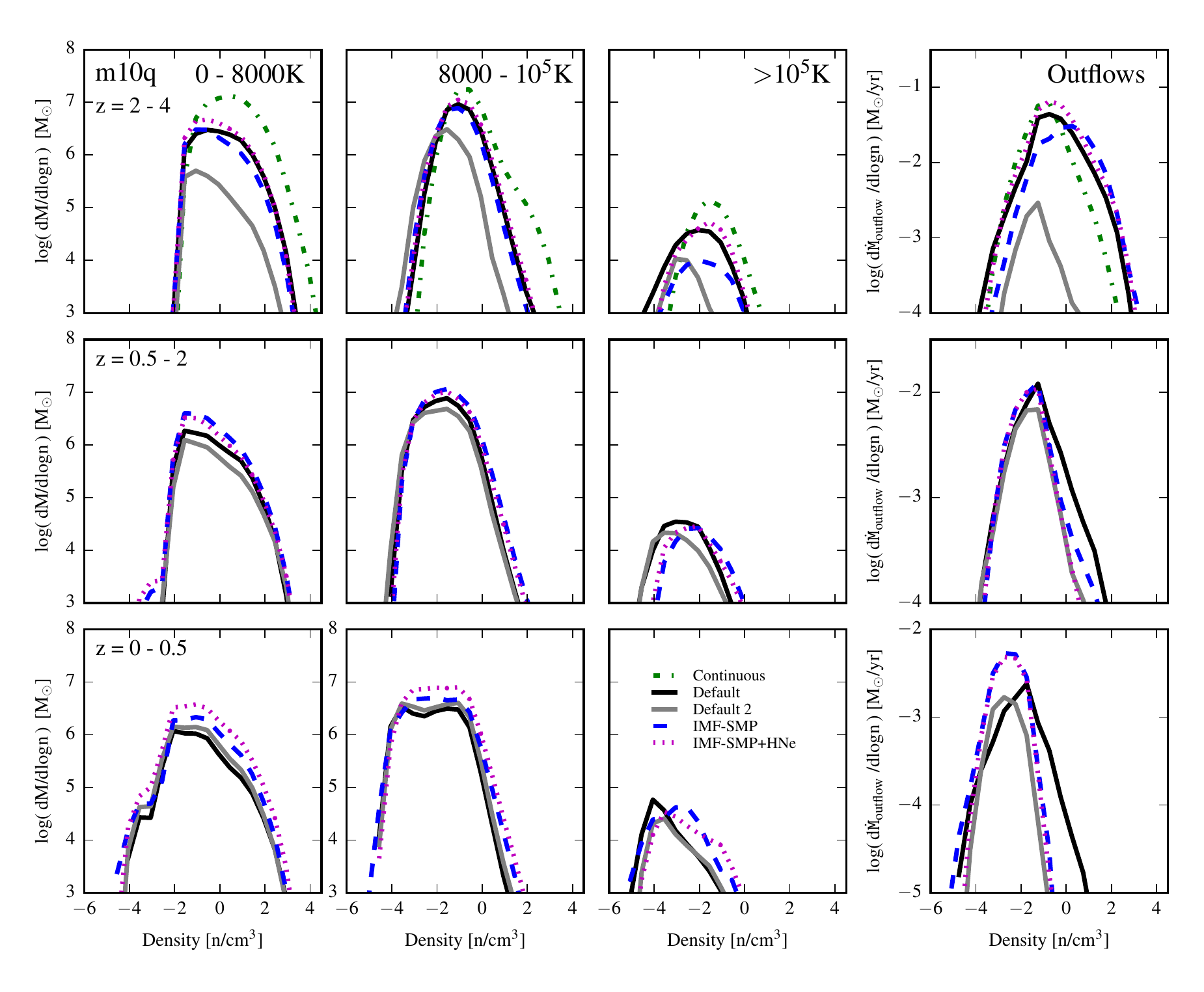}
\end{flushleft}
\label{fig:temperature_m10q}
\vspace{-0.7cm}
\caption{Gas density distributions for m10q. Rows show the properties at different redshifts; columns show phases including cold-neutral ({\em left}), warm-ionized ({\em middle left}), hot ({\em middle right}), and in outflows ({\em right}).  The ``Continuous'' run has more gas in all temperature bins, owing to less efficient feedback. Owing to the orders of magnitude higher stellar mass, it produces a significant outflow despite the feedback being effectively weaker.  \ksu{Given the stochastic difference between ``Default'' and ``Default 2'' runs, the effect of IMF sampling is not obvious.}}
\end{figure*}

\begin{figure*}
\begin{flushleft}
\centering
\includegraphics[width=18cm]{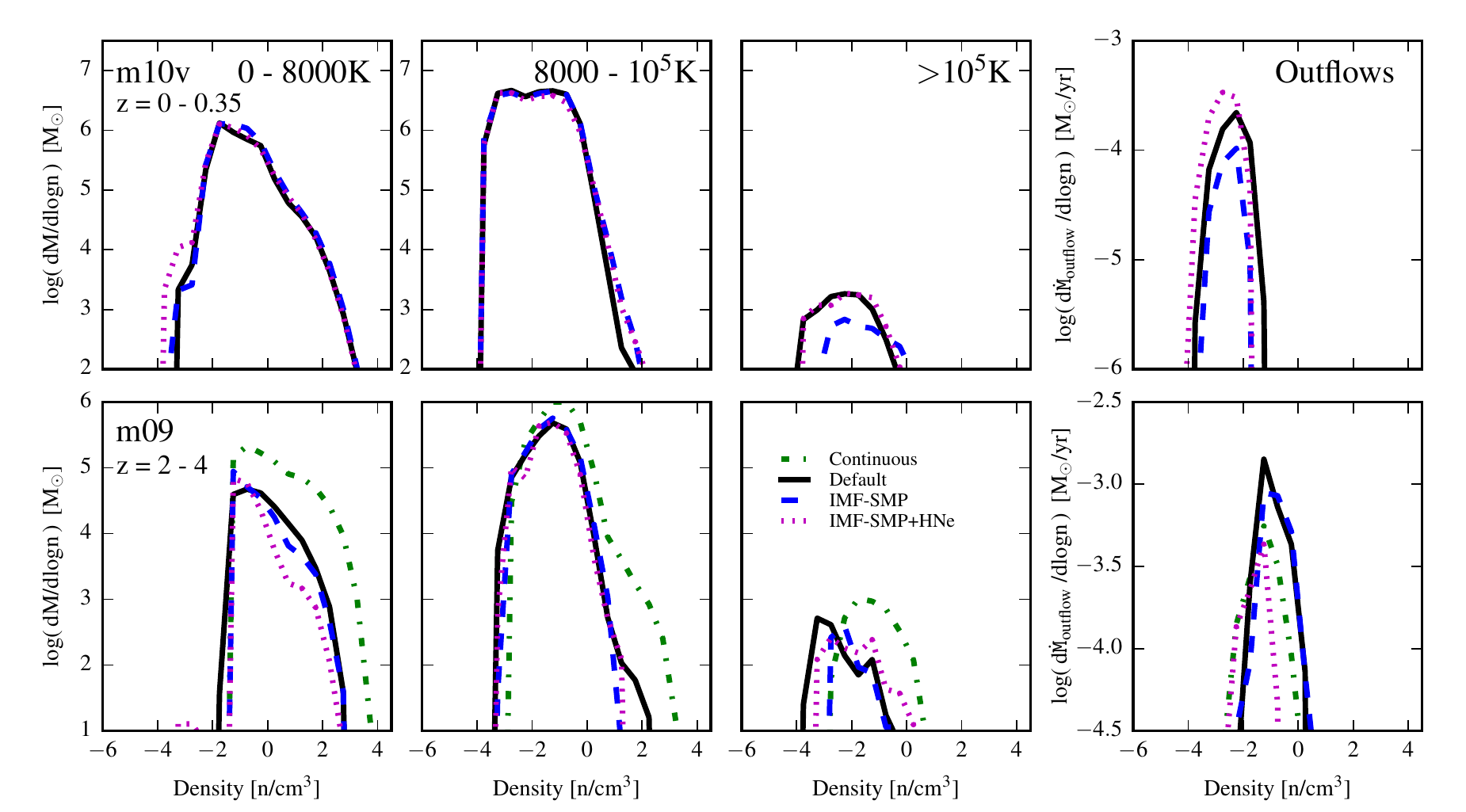}
\end{flushleft}
\label{fig:temperature_m10v09}
\vspace{-0.3cm}
\caption{Density distributions of outflows and gas in different phases as in \fref{fig:temperature_m10q}, but for m10v and m09. {\bf Top Row:} ``IMF-SMP+HNe'' run of m10v, from the time of the HNe ($z=0.31$) to $z=0$. {\bf Bottom Row:} m09 from $z=4$ to $z=2$.  The accretion rate of the ``Continuous'' run is higher, and therefore generates more cold and warm gas. HNe and IMF Sampling do not have large effects in these cases.  }
\end{figure*}

\fref{fig:temperature_m10q} and \fref{fig:temperature_m10v09} quantify the density distribution of gas particles in temperature bins of cold ($<8000$ K), warm (8000-$10^5$K) and hot ($>10^5$K) gas at various epochs. In the m10v case, since the ``IMF-SMP+HNe'' run is restarted from the ``IMF-SMP'' run at $z=0.31$ upon exploding a HNe and most of the star formation happens after that, only the low redshift  ($z=0-0.31$) results are shown. On the other hand, star formation in m09 ceases by $z\sim2$ and therefore only $z=2-4$ results are shown.

Again, the ``Continuous'' runs differ from the other runs most dramatically in all cases. All the runs with continuous SNe have higher total gas mass, especially in the cold and warm temperature bins. The total stellar mass is also orders of magnitude higher, which indicates that, without discretizing SNe, feedback is much less efficient and more gas can accretes onto the galaxy.

\ksu{The lack of cold gas in m10q ``Default 2'' run during the $z=2-4$ interval is consistent with its lower SFR in the same period. The lower SFR also results in less hot, intermediate density gas. Given the difference between m10q ``Default'' and ``Default 2'' runs,  the effect of IMF sampling on phase structure is not obvious.  IMF sampling does not appear to systematically alter the phase structure of the gas in m10v and m09 as well.} Since FIRE dwarf galaxies at this mass scale have relatively bursty star formation histories \citep{2016ApJ...820..131E,2017MNRAS.471.3547F, 2017MNRAS.466...88S, 2017arXiv170104824F}, IMF sampling is likely subdominant to bursts in establishing the phase structure of gas in these simulations.

In all cases, HNe do not alter the phase structure significantly. Whenever a HNe occurs, its effects only last for a few million years.

\subsection{Outflows}
\label{S:outflow}
The third row of \fref{fig:sfr} shows the outflow rate as a function of time in the simulations. The value shown is averaged over a 100 Myr period. To isolate ``outflows'', we simply take all gas within a thin layer from 0.08 to 0.1 $r_{vir}$ that has an outward radial velocity greater than $30 {\rm km\,s^{-1}}$ (comparable to the circular velocity in these dwarfs). The bottom row of \fref{fig:sfr} is the outflow mass-loading, defined as $\dot{M}_{\rm outflow}/\dot{M}_{\rm SFR}$, indicating the efficiency of stellar feedback at driving outflows. The plotted mass-loading is averaged over 500 Myr, to suppress stochastic effects. The density distributions of the outflows are shown in the fourth columns of \fref{fig:temperature_m10q} and \fref{fig:temperature_m10v09}.

The ``Continuous'' runs again demonstrate fundamental differences: despite having similar  outflow masses to the other runs, the star formation rate in the ``Continuous'' runs is an order of magnitude higher and the mass-loading is therefore much lower. This indicates that without discretizing SNe,  the ``smeared'' SNe energy injection is much less efficient  at accelerating gas into outflows. 

\ksu{The difference in outflows among the other m10q runs is consistent with the variation in star formation rates as the difference in outflow mass-loading is not significant during most of the time. This suggests that feedback efficiency in each run is similar on the average. Given the stochastic variance we see from the ``Default'' and ``Default 2'' runs, the effect of IMF sampling is again not obvious.}

%\ksu{On the other hand, the differences among runs are  visible only in the m10q cases,} in which both the ``IMF-SMP'' and ``IMF-SMP+HNe'' runs have higher outflow rates. These are largely a result of slightly higher SFRs in those runs as the difference in outflow mass-loading is not significant during most of the time, which suggest that difference of feedback efficiency in each run is also small on the average. The lower outflow mass-loading of the ``Default'' run at early times is mostly a consequence of its slightly earlier star formation episode, which suppresses star formation for $\sim 1$ Gyr after that. \ksu{Similar to the difference in star formation histories, the difference here is also stochastic.}

A peak of outflow can be seen just right after the manually-exploded HNe in the m10v and m09 cases. However, the long-term effects of HNe in these runs are, again, not obvious.

\subsection{Ionizing photon escape fractions}
To investigate the ionizing photon escape fractions, we follow the method in \cite{2015MNRAS.453..960M,2016MNRAS.459.3614M}. All the snapshots are processed by the 3 dimensional Monte Carlo radiative transfer (MCRT) code, basing on {\sc SEDONA} base  \citep{2006ApJ...651..366K}. For each snapshot, the intrinsic  photon budget $Q_{int}$ is calculated as the sum of the photon budget of each star particle  estimated through the BPASSv2 \citep{2016MNRAS.456..485S} model, which includes detailed binary evolution effects. Because the model stellar evolution  tracks exists only for certain metallicities, the input metallicity is assumed to be 0.001 (0.05 $Z_\odot$) \footnote{We use $Z_\odot=0.02$ \citep{1989GeCoA..53..197A}.}, which is roughly the averaged value in the simulations. We also assume $40\%$ of the metals are in dust phase with opacity $10^4 $cm$^2 $g$^{-1}$ \citep{1998ApJ...501..643D,2011MNRAS.418.1796F}. In the runs considering the effects of sampling the IMF, the photon budget from each star is scaled properly with its O-star number.
\vspace{1.5cm}

The MCRT code includes photo-ionization  \citep{1996ApJ...465..487V}, collisional ionization \citep{1968slf..book.....J}, and recombination \citep{1996ApJS..103..467V}. We run the calculation iteratively to reach converged results by assuming the gas in ionization equilibrium. The escape fraction is defined as the $Q_{esc}/Q_{int}$, where $Q_{esc}$ is the calculated number of escaped photons at approximately $R_{vir}$. Some examples of convergence test can be found in \cite{2015MNRAS.453..960M}.

\begin{figure*}
\begin{flushleft}
\centering
\includegraphics[width=18cm]{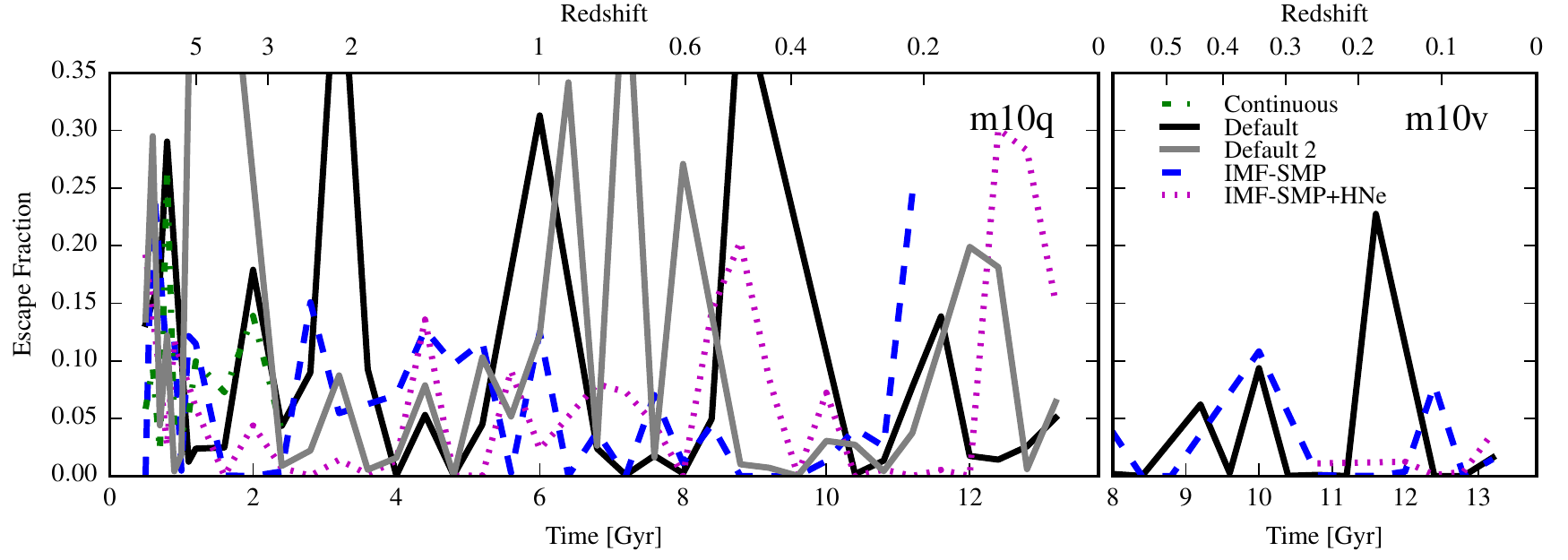}
\end{flushleft}
\label{fig:escape}
\vspace{-0.3cm}
\caption{Photon escape fractions ($Q_{esc}/Q_{int}$) for the m10q and m10v cases. No systematic effect from IMF Sampling, SNe discretization, or HNe is observed.}
\end{figure*}

\fref{fig:escape} shows the 400 Myr-averaged escape fraction for m10q and m10v runs. There are very few snapshots with young star particles ($<5$ Mry old, when most ionizing photons are emitted) in m09 and in m10v before $z=0.6$, so the results in those periods are poorly sampled and are therefore not shown. The photon escape fractions are highly variable during the simulated period, ranging from $\lesssim 0.001$ to 0.25, but no systematic effect from different models is observed.

The effects of IMF sampling on photon escape fractions are small. IMF sampling mainly affects the photon budget when there are O stars in the star particles. However, those stars are mostly deeply buried in dense GMCs from which  the photons rarely escape in any case.

\section{Discussion} \label{S:discussion}

\subsection{IMF Sampling effects}
%In m09 and m10v, we see no obvious effects from our IMF sampling model (in the properties we have analyzed). In the m10q case in \fref{fig:sfr}, the runs with IMF sampling not only form slightly more stars, but also have slightly smoother star formation histories. The default run, on the other hand, has a somewhat higher early SFR, leading to more gas expelled and a lower late-time SFR. However, this is due to stochastic or chaotic variations in the details and timing of different bursts, as such differences does not happen in the extra ``Default'' and ``IMF-SMP'' runs. 

We see no obvious effects from our IMF sampling model (in the properties we have analyzed). Our implementation of IMF sampling is based on a simple scaling of the local magnitude of feedback according to the number of massive O stars.  Those GMCs with higher O-star number can be destroyed more easily by feedback (both from SNe and ``pre-processing'' radiative feedback and OB-winds) and form fewer stars in their lifetime. On the other hand, in the regions  (periods) where (when) there are fewer O stars, the effects of feedback are weaker and therefore the gas accretion rate increases. %Given that  the accretion flows in dwarfs are highly concentrated in certain time intervals, and that the effect of boost/suppression of feedback is highly non-linear, it is possible that undershooting of O-star can have larger effect than overshooting on accretion.  As a result, the net effect can increase the total (time-integrated) accretion. The higher accreted gas mass can boost the later star formation rate, and, therefore, also the outflow rate.

In  larger halos (e.g. SMC-mass and larger), which form orders of magnitude more stars and have much deeper potential wells, phenomena such as galactic winds result from the collective effects of many stars. Hence, the local variation of O-star number will be less significant.

\ksu{On the other hand, in the halos where many fewer stars are formed (e.g. dwarfs like m09, m10v or m10q), the amount of gas in the close neighborhood of young stars is reduced} and a single SNe (which is already discretized in these simulations) has a large feedback effect regardless of whether or not other SNe explode nearby. As a result, the spatial and time variation of the local magnitude of feedback is already large, and IMF sampling may be a secondary effect compared to strong stellar clustering. 

%It is plausible, then, that m10q may be in the sweet spot where IMF sampling affects galaxy evolution. In the region (period) where (when) the O-star number is lower, gas accretes more efficiently, resulting in higher gas masses, which later boost the star formation rate and outflows.

\ksu{It is also worth noting that IMF sampling does not statistically change the spatial and time distribution of SNe events (primarily determined by the distribution of star formation events, which trace the dense, self-gravitating ISM gas), other than linking the strength in each feedback channel to the local O-star number. In other words, it does not on average increase SNe rate, and nor does it make SNe more or less clustered. 
%Instead, it links the strength in each feedback channel to the O-star number.
}

\ksu{In the runs with IMF sampling, the number of O stars is drawn from a Poisson distribution with mean and variance equal to the average number of O stars. Regardless of the random numbers drawn, most O stars will explode as SNe within 30 Myr. As a result, the statistical properties of SNe are roughly the same as with the default physics. %, where the SNe in each time step follows a binomial distribution given the time step is small. 
An important difference in runs with IMF sampling is that star particles with higher O-star numbers will not only have more SNe but also generate more powerful stellar wind and radiative feedback (instead of IMF-averaged). In other words, the modified SNe feedback is synchronized with the other feedback channels. Although this may further boosts the total feedback strength in different regions beyond merely the variation in SNe, such boost is probably modest if SNe are the dominant feedback mechanism, which is often the case in dwarfs like m10q, m10v and m09.}

\subsection{Ineffective HNe feedback}

By construction, in m10q and m10v, the {\em total} injected HNe energy in the simulation period is sub-dominant at $\sim 10\%$ of the integrated SNe energy. In the m09 run, the energy injected by the HN is comparable to total energy injected by SNe throughout the simulation, because the galaxy has so few stars. %In all cases, however we do not see qualitatively different galaxy evolution with HNe,

However, just one HN is equivalent to 100 overlapping regular SNe. As such, we see a single HN can eject a large fraction of the core ISM in these low-mass dwarfs, and successfully suppress star formation for $\sim 1$ Gyr. Eventually, the gas recycles and begins the next cycle of star formation - it is worth noting that even $\sim 10^{53}$ erg can only accelerate $\lesssim 10^6 M_{\odot}$ of gas to speeds of order the escape velocity in these systems. However, in our simulations, the star formation in such low-mass dwarfs is highly bursty, and highly concentrated in some time intervals. In m10q or m10v, $\gtrsim 10^4 M_\odot$ stars can form in certain 100-million-year periods. In m09, although only $\sim 10^4 M_\odot$ form in the simulation, roughly half of that forms in the largest star burst. As a result, although HNe are very extreme versions of SNe, $\sim 100$ overlapping SNe do happen in the simulations occasionally, and have similar effects. Therefore, including HNe in the simulations does not appreciably alter galaxy evolution, in a statistical sense, compared to ``normal'' clustered and bursty star formation.

\section{Conclusion}
\label{S:conclusions}
In this study, we have investigated the effects of various discrete processes in stellar feedback, including SNe, HNe and IMF sampling on the formation and evulution of dwarf galaxies with stellar masses in the range of $\sim 10^4-3\times 10^6M_\odot$. We summarize our conclusions below.
\begin{itemize}
\item Discretizing SNe injection is crucial. Treating SNe as continuous energy/momentum sources with time-averaged rates (instead of individual events) smears the energy in time and space, which allows it to radiate away far too efficiently. This severely the exacerbates the so-called ``overcooling problem''. As a result, feedback is effectively weaker, making galaxies accrete more gas and form orders of magnitude more stars.
\item  \ksu{ Given the purely stochastic simulation variations between m10q``Default'' and ``Default 2'' runs, the effects of IMF sampling are not obvious. IMF sampling also has no obvious effect on the smaller and burstier galaxies (m10v or m09).} %In the m10q cases, runs with IMF sampling accrete slightly more gas and have slightly higher SFRs. Although they generate more outflows, the wind mass-loading is not higher, as the result of higher SFRs (i.e. feedback is similarly ``efficient''). 
\item HNe and IMF sampling effects as approximated here do not systematically affect the photon escape fraction at an appreciable level in our analysis.
\item  The effects of HNe are not obvious in the investigated cases. While dramatic as individual events when they occur, and capable of ejecting gas and shutting down SF temporarily (for up to $\sim1$ Gyr) in ultra-faint dwarfs, they resemble overlapping SNe from star clusters, so do not qualitatively change galaxy evolution in an aggregate, statistical sense. Since the ISM gas ejected by HNe is mostly recycled after $\sim 1$ Gyr, it should be possible to observe HNe yields in next-generation stars in faint dwarfs.
\end{itemize}

We caution that the toy model here for IMF sampling only scales feedback strength with some ``O-star number''. More accurately, one should drawn a mass spectrum from the IMF, and some properties (e.g. photo-ionization) will be more strongly sensitive to the most massive stars. Of course, these will also produce distinct yields when they explode.  Moreover, HNe should  have different enrichment properties. HNe rate is also connected with the IMF, which could be redshift dependent. At high redshift, the HNe event rate can be 10 times higher than in low redshift \citep{2012Natur.491..228C}, which can possibly further change the halo mass at reionisation, and therefore also the post-reionisation accretion. These aspects are left for future work. Besides the discreteness in feedback processes investigated in the current study, there are other processes that could be interesting and can be crucial in galaxy evolution. For instance, SNe injection should also affect the cosmic-ray energy budget, which is not included in the current feedback model. Detailed examination of these processes will also be left for future work.

%\vspace{-0.7cm}
\acknowledgments

Support for PFH was provided by an Alfred P. Sloan Research Fellowship, NASA ATP Grant NNX14AH35G, and NSF Collaborative Research Grant \#1411920 and CAREER grant \#1455342. The Flatiron Institute is supported by the Simons Foundation. Numerical calculations were run on the Caltech compute cluster ``Zwicky'' (NSF MRI award \#PHY-0960291) and allocation TG-AST130039 granted by the Extreme Science and Engineering Discovery Environment (XSEDE) supported by the NSF. D. Kere\v{s} was supported by NSF grants AST-1412153 and  AST-1715101 and the Cottrell Scholar Award from the Research Corporation for Science Advancement. CAFG was supported by NSF through grants AST-1412836, AST-1517491, AST-1715216, and CAREER award AST-1652522, and by NASA through grant NNX15AB22G. MBK acknowledges support from NSF grant AST-1517226 and from NASA grants NNX17AG29G and HST-AR-13888, HST-GO-14191, HST-AR-14282, and HST-AR-14554 from the Space Telescope Science Institute, which is operated by AURA, Inc., under NASA contract NAS5-26555. M. Orr is supported by the NSF GFRP under grant No 1144469.
\\

\footnotesize{
\bibliographystyle{mn2e}
\bibliography{mybibs}
}

\label{lastpage}

\end{document}